\documentclass[fleqn,11pt]{wlscirep}
\usepackage[utf8]{inputenc}
\usepackage[T1]{fontenc}
\usepackage{amsmath,bm}
\title{Non-destructive mapping of stress, strain and stiffness of thin elastically deformed materials}

\author[1]{Guo-Yang Li}
\author[2]{Artur L. Gower}
\author[3,4,*]{Michel Destrade}
\author[1,*]{Seok-Hyun Yun}

\affil[1]{Harvard Medical School and Wellman Center for Photomedicine, Massachusetts General Hospital, Boston, MA 02139, USA.}
\affil[2]{Department of Mechanical Engineering, University of Sheffield, Sheffield, United Kingdom.}
\affil[3]{School of Mathematical and Statistical Sciences, NUI Galway, Galway, Ireland.}
\affil[4]{Key Laboratory of Soft Machines and Smart Devices of Zhejiang Province, Department of Engineering Mechanics, Zhejiang University, Hangzhou 310027, PR China.}
\affil[*]{Corresponding authors: michel.destrade@nuigalway.ie (M.D.); syun@hms.harvard.edu (S.H.Y.).}

\begin{abstract}
Knowing the stress within a soft material is of fundamental interest to basic research and practical applications, such as soft matter devices, biomaterial engineering, and medical sciences. However, it is challenging to measure stress fields \textit{in situ} in a non-invasive way. It becomes even more difficult if the mechanical properties of the material are unknown or altered by the stress. Here we present a robust non-destructive technique capable of measuring \textit{in situ} stress and strain in elastically deformed thin films without the need to know their material properties. The technique is based on measuring elastic wave speeds, and then using a universal dispersion curve we derived for Lamb wave to predict the local stress and strain. Using optical coherence tomography, we experimentally verified the method for a rubber sheet, a cling film, and the leather skin of a musical instrument.

\vspace{5 mm}
\textbf{keywords}: Mechanical stress $|$ Lamb wave $|$ Acoustoelasticity $|$ Optical coherence elastography $|$ Soft matter
\end{abstract}

\begin{document}
\flushbottom
\maketitle
\thispagestyle{empty}


\newpage
\section*{Introduction}
\noindent
Soft thin films hanging in the air or  confined in fluids are ubiquitous in our daily lives as well as in natural and engineering systems.
Examples include cling film packaging food, the eardrum and the diaphragm in our body, and various elastic sheets, membranes, vesicles and bands holding structures together.
They are typically under external and internal stress, and it is often desirable to know the stress level to be able to understand the environment they are exposed to or interacting with, or to monitor the changes in and health of the materials.
However, \textit{in situ} non-invasive measurements of the stress are challenging.
This is even more challenging if the mechanical properties of the material are unknown and, furthermore, if the original configuration of the material is unknown, which precludes straightforward measurement of strain~\cite{Gomez-Gonzalez2020,Schajer2009}.

Various techniques have been devised to measure in-plane stresses~\cite{butt2013physics}.
The choice of technique depends on the  material  type (solid/liquid/type of molecules) and also on the length scale. Essentially all techniques so far rely on the knowledge of the elastic moduli of the material or some specific expected behaviour of the material, which limits their application to known or specific materials and structures.
For example, in the Langmuir–Blodgett trough~\cite{erbil2006surface}, a workhorse of membrane biophysics, surface tension is estimated by measuring the amount of force required to insert a Wilhelmy plate into a given membrane.
However, this force depends on the nature of the surface tension and its accuracy has been questioned for solid-like membranes~\cite{aumaitre2011measurement}.
Conventional ultrasound methods also require the elastic moduli and acousto-elastic parameters of the materials to predict the stress~\cite{shi2013,li2020ultrasonic}.

Here we describe a technique that allows the stress field to be determined in soft thin films even without \textit{a priori} knowledge of the material properties or applied strain.
The technique uses Lamb elastic waves propagating in the film~\cite{Lanoy2020,Thelen2021} followed by a simple algorithm to determine the stress from measured wave speeds.
In this work, we use optical coherence tomography (OCT) to visualise the elastic waves and measure their propagation speeds in an audible frequency range.
This range (1-20 kHz) is well suited for soft materials with thickness ranging from sub-micron to a few hundreds of micrometers.

\section*{Results}
\subsection*{Theoretical foundation}
\begin{figure*}[h!]
    \centering
    \includegraphics[width=.9\textwidth]{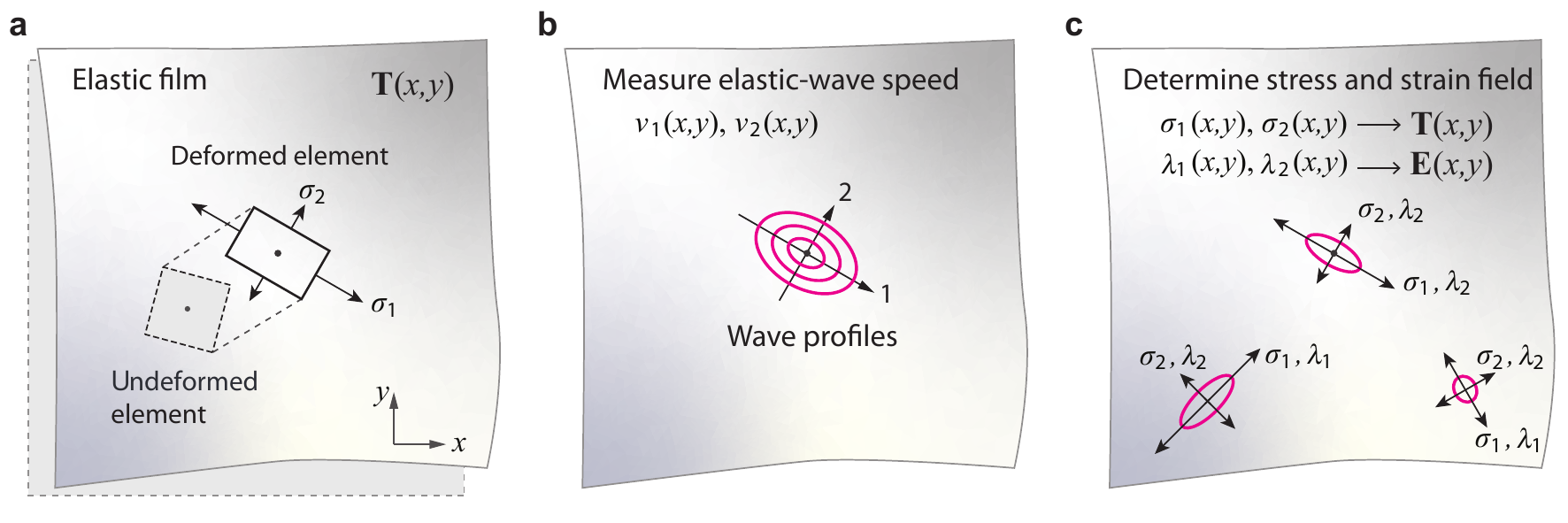}
    \caption{\textbf{The principle of measuring stress and strain via acousto-elastic mapping.}
    \textbf{a}, Thin films elastically deformed by the stress field $\mathbf{T}$ from an (unknown) undeformed configuration.
    A representative dashed square is deformed into a rectangle by the principal stresses $\sigma_1$ and $\sigma_2$ at ($x$, $y$).
    \textbf{b}, Wave profiles in the stressed membrane showing anisotropic wave speeds. The wave speeds along the principal directions, $v_1$ and $v_2$, are used to recover the principal stresses and stretches at ($x$, $y$).
    \textbf{c}, The principal stresses and stretches measured at different locations give the stress $\bm{\sigma}$ and strain $\mathbf{E}$ map of the membrane.}
    \label{fig:figure1}
\end{figure*}

Figure~\ref{fig:figure1} illustrates the general principle for a film under an arbitrary stress field \textbf{T} which deformed it elastically.
The local wave speed is affected not only by the stiffness of the material but also by the direction and magnitude of the local stress (Fig.~\ref{fig:figure1}b)~\cite{Hughes1953PR}.
The two principal stresses $\sigma_1$ and $\sigma_2$ and stretch ratios $\lambda_1$ and $\lambda_2$, at a location $(x,y)$ are related to the in-plane stress (Cauchy stress) $\bm{\sigma}$ and strain (Green-Lagrange strain) $\mathbf{E}$ at the location (Fig.~\ref{fig:figure1}c).
Mathematically, $\sigma_1$ and $\sigma_2$ are obtained by diagonalising the stress tensor.

\begin{figure*}[t!]
    \centering
    \includegraphics[width=.98\textwidth]{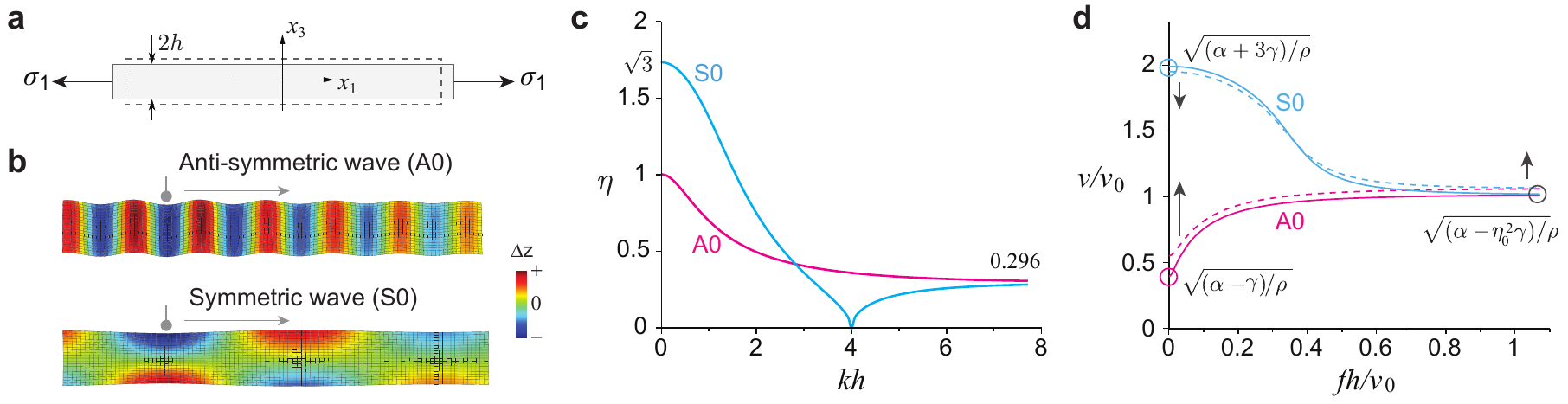}
    \caption{\textbf{Acousto-elastic effect for Lamb waves.}
    \textbf{a}, A thin structure under tension.
    \textbf{b}, Modal shapes of the fundamental anti-symmetric (A0) and symmetric (S0) Lamb waves.
    \textbf{c}, Global curves for the A0 and S0 modes.
    \textbf{d}, Representative dispersion curves of the A0 and S0 modes in an incompressible material with shear modulus $\mu=1.0$ kPa, Landau constant of third-order elasticity $A=-3.2$ kPa and mass density $\rho = 1$ g/cm$^3$, subject to two different strains of 5\% (full lines) and 10\% (dashed lines) that are applied to the wave progagation directions. The $\alpha$, $\gamma$, and $v$ are equal to $\alpha_i$, $\gamma_i$, and $v_i$, depending on direction of propagation. The circles indicate three asymptotic values of the dispersion curves, and arrows indicate their direction of change with tension, see Methods for more details.}
    \label{fig:figure2}
\end{figure*}

To develop our algorithm, we consider a deformed element (Fig.~\ref{fig:figure2}a) with  local coordinates $(x_1,x_2,x_3)$, where $-h\leq x_3 \leq h$, $2h$ being the thickness.
The film is subject to the in-plane stresses $\sigma_1$ and $\sigma_2$ along the $x_1$ and $x_2$ axes, respectively, and the out-of-plane stress $\sigma_3$ along the $x_3$ axis ($\sigma_3 \simeq 0$ in thin films).
An infinitesimal elastic wave polarised in the $x_i - x_3$ plane and propagating along the $x_i$ axis ($i$ = 1 or 2) is described with the mechanical displacement field $\mathbf{u} = \mathbf{u}_0 \exp{(-s \, k \,x_3)} \exp{[ik(x_i - v_i \,t)]}$, where $\mathbf{u}_0$ is the amplitude vector, $k$ is the wavenumber, $s$ is the attenuation factor, $t$ is time, and $v_i$ is the speed.
The governing wave equation is~\cite{Ogden2007} $\mathcal A^0_{pjqk}\partial^2 u_k/ \partial x_p \partial x_q - \partial \hat{\Bar{p}} / \partial x_j = \rho \partial^2 u_j / \partial t^2$.
Here $\mathcal A^0_{pjqk}$ is the Eulerian elasticity tensor, which contains the effect of the stress via the strain energy, and the second term $\hat{\Bar{p}}$ denotes the increment of the Lagrange multiplier $\Bar{p}$, due to the constraint of incompressibility, and $\rho$ is the mass density.
Eliminating $\hat{\Bar{p}}$ with incompressibility ($\lambda_1 \lambda_2 \lambda_3 = 1$, $\nabla \cdot \mathbf{u} = 0$), the wave equation is then reduced to~\cite{ogden1993effect}: $\gamma_i  \, s^4 - (2\beta_i  - \rho v_i^2) \, s^2 + \alpha_i  - \rho v_i^2 = 0$,
where $\alpha_i  = \mathcal A^0_{i3i3}$, $2\beta_i  = \mathcal A^0_{iiii} + \mathcal A^0_{3333} - 2\mathcal A^0_{ii33} - 2 \mathcal A^0_{3ii3}$, and $\gamma_i = \mathcal A^0_{3i3i}$.
The stress-free boundary condition ($\sigma_3 = 0$ at $x_3 = \pm h$) of the thin film structure gives the complete dispersion equation of the Lamb waves (see Methods).

The instantaneous elastic moduli, $\alpha_i$ and $\gamma_i$ in elastically deformed materials are directly related to the principal stresses and stretches through the identities (see Methods)
\begin{equation} \label{eq:alphagamma}
\sigma_i = \alpha_i  - \gamma_i;  \quad \lambda_i^2 / \lambda_3^2 = \alpha_i  / \gamma_i.
\end{equation}
For isotropic materials, with up to third-order elasticity~\cite{destrade2010third}, we have that $2\beta_i = \alpha_i + \gamma_i$.
This identity simplifies the dispersion equation to:
\setlength{\belowdisplayskip}{6pt} \setlength{\belowdisplayshortskip}{6pt}
\setlength{\abovedisplayskip}{6pt} \setlength{\abovedisplayshortskip}{6pt}
\begin{equation} \label{eq:dispersion}
4 \, \eta \, \tanh{(\eta kh)} / \tanh{(kh)} = (1+ \eta^2)^2
\end{equation}
for the anti-symmetric (A) modes, and
$4 \, \eta \, \tanh{(kh}) / \tanh{(\eta kh)}  = (1+ \eta^2)^2$
for the symmetric (S) modes, where $\eta \equiv  \sqrt{|\alpha_i - \rho v_i^2| / \gamma_i}$.
The fundamental A0 mode is a flexural bending wave, and the fundamental S0 mode has a dilatational, out-of-plane displacement ($u_3$) profile (Fig.~\ref{fig:figure2}b).

Our algorithm to determine the in-plane stress and strain is as follows. First, note that $\eta$ is uniquely related to $kh$ via the dispersion equation, so that $\eta = \eta(kh)$.
This relationship for the two modes is displayed in Fig.~\ref{fig:figure2}c.
Second, we experimentally measure $k$ or $v_i$ at certain frequencies $f$.
Note that $v_i = 2 \pi f / k$. Then, $\eta(kh)$ is determined using the dispersion relationship.
When $\eta$ is measured at two different $f$'s, we can determine $\alpha_i$ and $\gamma_i$ from the two values of $\eta$.
Although two frequency data are sufficient in principle, measurements over multiple frequencies, followed by a least square fit, lead to a more accurate predict of $\alpha_i$ and $\gamma_i$.
Finally, the principal stress, $\sigma_i$, is determined from the first equation in Eq.~\eqref{eq:alphagamma}. The principal strain, $E_i = 0.5 (\lambda_i^2 - 1)$, is determined from the second equation in Eq.~\eqref{eq:alphagamma}. With the stress and strain, the Young's modulus (for small strain) can be determined as $(\sigma_i - 0.5\sigma_j) / E_i$.

For insight, let us consider an intrinsically isotropic nonlinear material under uni-axial tension.
Figure~\ref{fig:figure2}d shows the dispersion curves of the wave speeds normalized to the bulk shear wave speed, $v_0 = \sqrt{\mu/\rho}$ of the material in the undeformed state, where $\mu$ is the shear modulus (unknown \textit{a priori} in experiments), are plotted as a function of normalized frequency, $f h/v_0$  for two different extension values of 5\% (full lines) and 10\% (dashed lines), respectively.
For small  deformation ($\lambda_i$ close to 1), the A0 and S0 wave speeds along the $x_1$ axis in the limit of low frequency are $v_{A0} (0) = \sqrt{\sigma_1 / \rho}$ and $v_{S0}(0) = 2\sqrt{\mu /\rho} - \frac{1}{6}\sqrt{(\sigma_1 + \sigma_2)^2 / (\rho \mu)}$ (see Methods).
In this case ($\mu \gg \sigma_i$), the A0 wave speed has much higher sensitivity than the S0 wave to the principal stress along the propagation direction.
A more rigorous sensitivity analysis (Supplementary note 1) supports this conclusion for larger deformations.
In our experiments, we exclusively used A0 waves.

Although our algorithm does not require the knowledge of $\mu$ to measure stress, the knowledge of thickness $2h$ makes the algorithm more robust in determining the stretch and is needed to measure the elastic modulus of the material at the deformed state.
The thickness information may be obtained by imaging.
For most elastomers and biological tissues,  $\rho \approx$ 0.9 - 1.1 g/cm$^3$ and $\mu$ is in a range of 1 kPa to 1 GPa.
Then, $v_0 \approx$ 1 to 1000 m/s. Measurement over an audible range, $f$ = 1 - 20 kHz, allows us to measure samples with $h$ $\sim$ 1 to 500 $\mu$m.

\begin{figure*}[ht!]
    \centering
    \includegraphics[width=.95\textwidth]{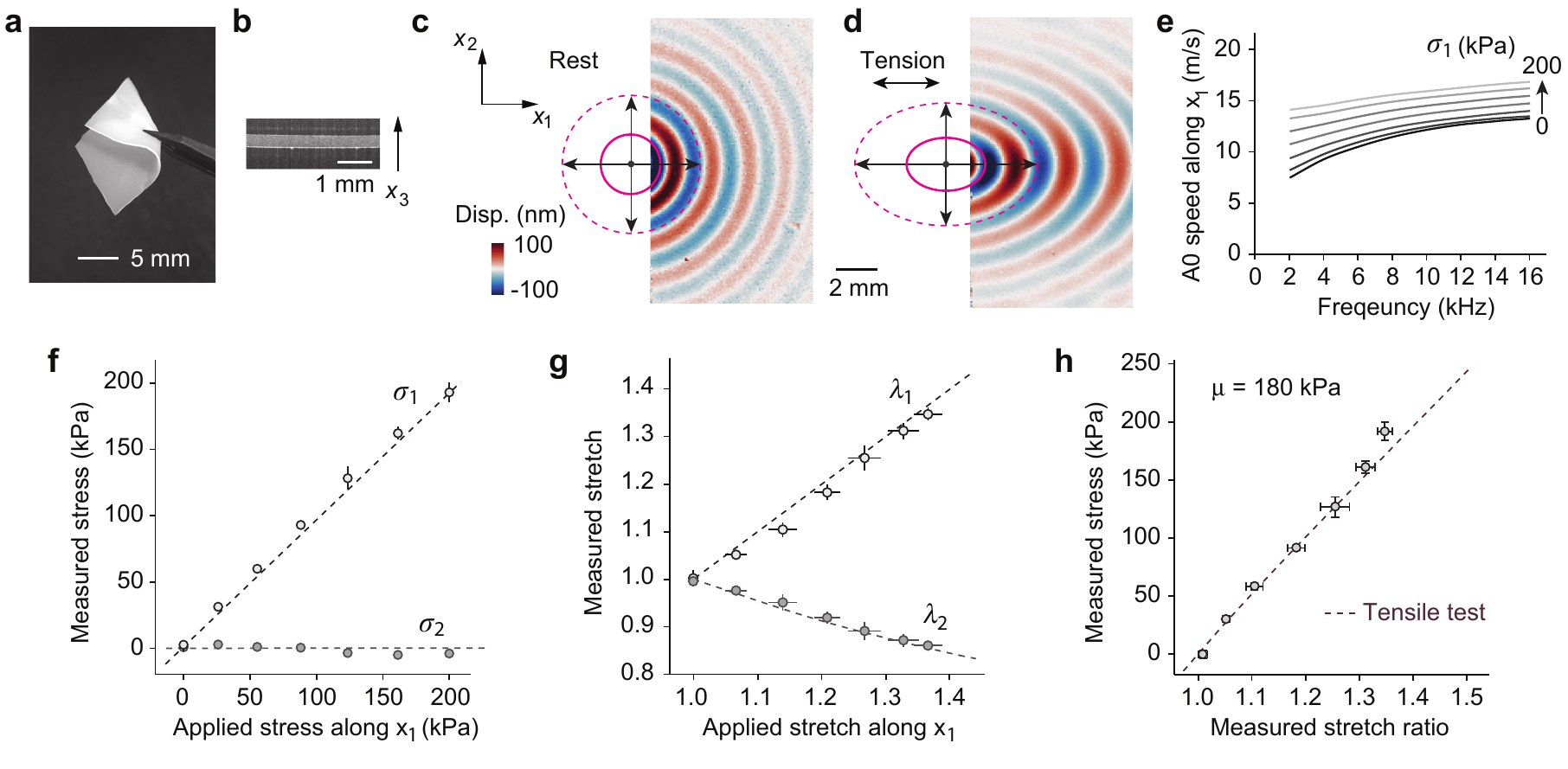}
    \caption{\textbf{Imaging and analysing elastic waves in a stretched rubber sheet.}
    \textbf{a}, Photograph of our rubber sheet.
    \textbf{b}, OCT cross-sectional image.
    \textbf{c-d}, Wave profiles measured by OCT when the film is (\textbf{c}) stress-free and (\textbf{d}) subject to a uniaxial stress $\sigma_1 \approx 200$ kPa. The wavefronts becomes elliptical when stretched.
    \textbf{e}, Dispersion relations of the $A_0$ mode obtained at different levels of stress. Each curve is an average over six measurements (for clarity the error bar is not shown here).
    \textbf{f}, Comparison between the measured and applied stress values. For reference, the dashed lines show where the measured stresses are equal to the applied stresses.
    \textbf{g}, Comparison between the measured and applied stretch values. Dashed lines, $1$ and $-1/2$ power laws of the applied stretch for reference.
    \textbf{h}, The stress-strain curve from the measured data. Dashed line, stress-strain curve measured by standard tensile test.}
    \label{fig:figure3}
\end{figure*}

\subsection*{Experimental validation}
To validate the method, we devised an experimental setup (Supplementary Fig. S3) based on a home-built, swept-source OCT system~\cite{Antoine2019OE}.
The first sample used was a rubber sheet (Fig.~\ref{fig:figure3}a).
We applied uniaxial tension to the sample using weights (see Methods).
The thickness of the film measured by OCT (Fig.~\ref{fig:figure3}b) decreased from 500 to 430 $\mu$m
(Supplementary Fig. S4).
Figures~\ref{fig:figure3}c,d show the wave profiles in the unstressed and fully stressed states (0 and 6 20-g weights, respectively) at $f$ = 6 kHz (see Supplementary Movies 1 and 2).
In the unstressed state the wave propagates at the same speed in all directions, and creates a circular profile with $v_1 = v_2$. The applied stress, on the other hand, induces anisotropy for the speed, which creates an elliptical profile, with the speed $v_1$  along the tensile $x_1$ axis being larger than the speed $v_2$ along the compressive $x_2$ axis.
Figure~\ref{fig:figure3}e shows the dispersion curves of the A0 mode at different stress levels from 0 to 200 kPa.
We also measured the dispersion of the A0 wave along the $x_2$ axis (Supplementary Fig. S7).

The measured wave speeds $v_1$ and $v_2$ fit very well into the dispersion relation written in terms of $\eta$ (Supplementary Fig. S6), from which we determined $\alpha_1$ and $\gamma_1$, and $\alpha_2$ and $\gamma_2$.
Using Eq.~\eqref{eq:alphagamma}, we then obtained the stress and strain along the $x_1$ and $x_2$ axis.
Figure~\ref{fig:figure3}f shows the measured and actual values of the two principal stresses with a good agreement, with errors around 5\%.
Figure~\ref{fig:figure3}g shows the measured and actual stretch ratios again in a good agreement with errors less than 3\%.
The measured stress-strain curve (Fig.~\ref{fig:figure3}h) gives a value of $\mu \approx 177.8$ kPa, which agrees well with that obtained by an independent standard tensile test (Supplementary Fig. S8).

\begin{figure*}[bh!]
    \centering
    \includegraphics[width=.8\textwidth]{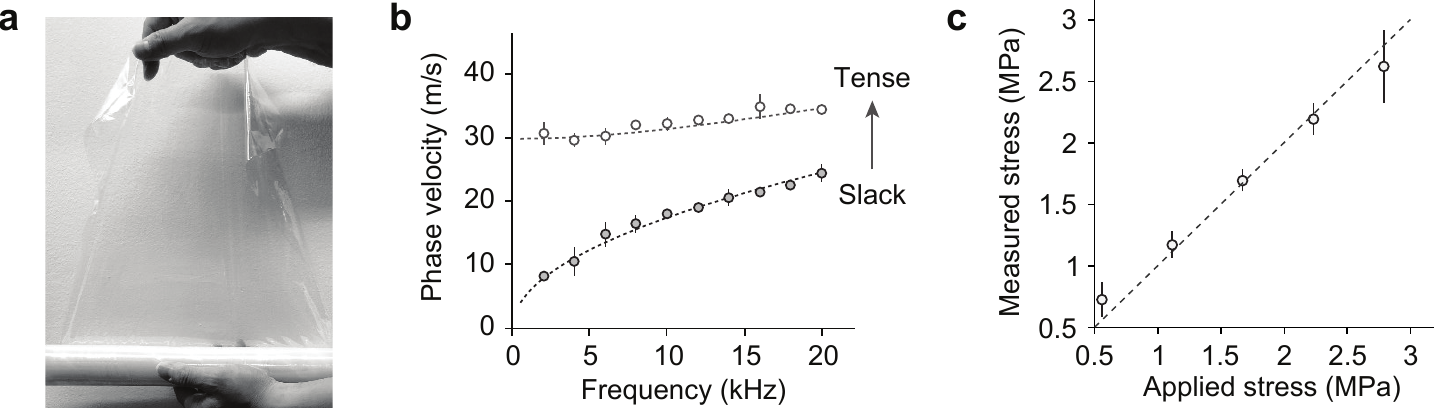}
    \caption{\textbf{Measuring mechanical stresses in a cling film.}
    \textbf{a}, Photograph of the cling film.
    \textbf{b}, Dispersion curves for a cling film in slack and tense states, respectively. Markers, experiment. Dashed curves, theory. The Young's modulus of the cling film fitted from the unstressed dispersion curve is $\sim 1170$ MPa.
    \textbf{c}, Comparison between the applied and measured stress. Dashed line, $45^\circ$ line for reference.
    }
    \label{fig:figure4}
\end{figure*}

Next, we tested the technique for a stretched plastic wrap, a.k.a. \emph{cling film} (Fig.~\ref{fig:figure4}a) made of polyethylene with a thickness of $\sim12$ $\mu$m.
Using a similar experimental setup, we applied a uniaxial stress to the cling film and measured the dispersion relations at the unstressed and stressed states (Fig.~\ref{fig:figure4}b). For the ultra-thin structure, the asymptotic speed in the low frequency limit provides $\sigma_i$ directly.
We measured the low-frequency wave speed $v_1(0)$ in the stretched condition to be $\sim30.3$ m/s. Using $\rho$ = 930 kg/m$^3$ we obtain $\sigma_1 \approx 0.84$ MPa. 
By Taylor expansion of Eq.~\eqref{eq:dispersion}, we find $\rho v_1^2 = \sigma_1 + (-\sigma_1 /3 + 4E/9) (kh)^2 + \mathcal{O}((kh)^4)$, where $E$ is the Young's modulus in the stretched condition.
By curve fitting the measured dispersion curve (Fig.~\ref{fig:figure4}b), we find $E \approx 1069$ MPa. This is slightly lower than the Young's modulus of $\sim 1170$ MPa in the unstressed condition. 
Experiments performed at different stretching force showed good agreements between the measured and applied stress values (Fig.~\ref{fig:figure4}c).

Finally we applied our technique to measure the stress in the drumhead of a musical instrument called the \emph{bodhr\'an drum}, a traditional Irish drum made with goat skin.
If the skin is too taut, the pitch is higher than expected, so players are advised to sprinkle and spread some water on the skin from the inside just before performance.
Conversely, if the skin is too loose, bodhr\'an players rub their palm on the outside of the skin to make it dry and tighten the skin to correct the pitch.

\begin{figure*}[h!]
    \centering
    \includegraphics[width=.8\textwidth]{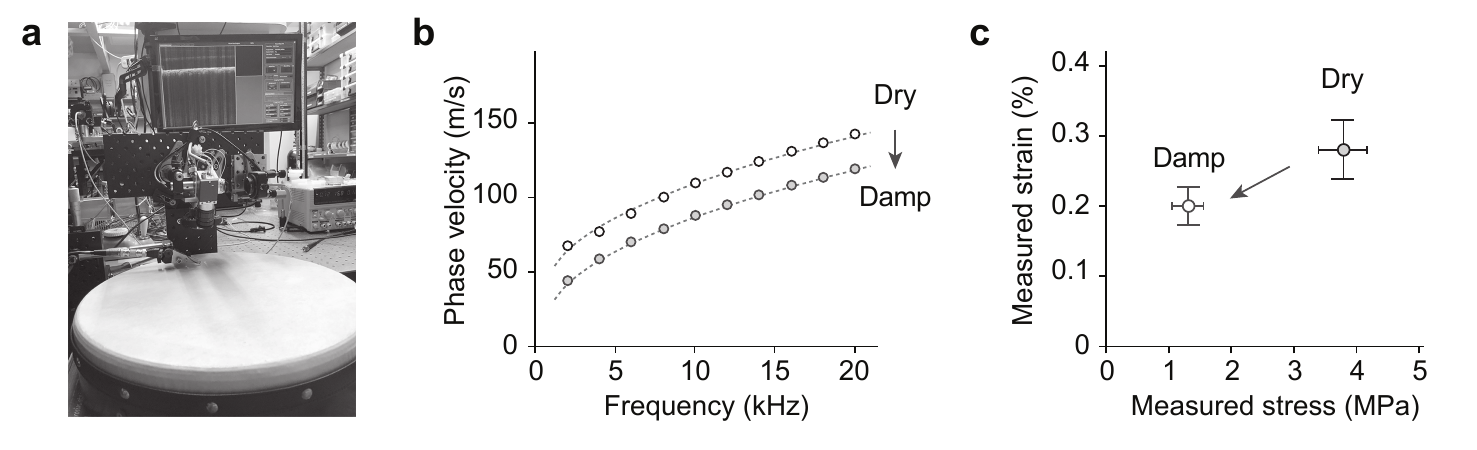}
    \caption{\textbf{Measuring mechanical stresses in a drum head.}
    \textbf{a}, Photograph of the experimental setup. The size of the \emph{bodhr\'an} drumhead is 16 inch. A real-time OCT image is displayed in the monitor.
    \textbf{b}, Dispersion relations measured in the dry and damp states. Markers, experiment. Dashed curves, theory.
    \textbf{c}, Stresses and strains measured in the dry and damp states.}
    \label{fig:figure5}
\end{figure*}

The thickness of the skin was measured to be $360 \pm 30$ $\mu$m.
We performed \emph{in situ} measurements on the drumhead (Fig.~\ref{fig:figure5}a) at normal (dry) and hydrated conditions of the skin. In the dry condition, the fundamental resonance frequency of the instrument was 84 Hz, and it was decreased to 36 Hz after hydration (Supplementary Fig. S9).
The goat skin is intrinsically anisotropic, but our experiments revealed almost circular wave profiles (Supplementary Movies 3 and 4).
This is well explained by the large radial stress in the drum, which stretches collagen fibres along the stress field \cite{deroy2017non}.
The applied strain is transversely isotropic (equi-biaxial in the radial/circumferential directions) and thus, for all intends and purposes, acoustic wave propagation is isotropic in the drum plane.
Figure~\ref{fig:figure5}b shows the dispersion relations of the skin in the dry and damp states, obtained at a region in the drum head.
We determined from the experimental data that the amount of radial stretch in the dry skin is 0.28\% and that humidification relaxed it to 0.20\% (Fig.~\ref{fig:figure5}c).
The corresponding stress is changed from 3.79 MPa (dry) to 1.31 MPa (damp).
Noting that the strain is small, we find the Young's modulus of the skin to be $\sim$ 680 MPa for dry skin and $\sim$ 330 MPa for the moisturized skin.

The result reveals how much the humidification changed the stiffness and consequently the tension of the skin. Hydration alters the resonance frequencies or pitch of the instrument, which is proportional to $\sqrt{\sigma_1 / \rho}$. The 65\% reduction in tension (and an increase of density) predicts $\sim$ 46\% decrease in resonance frequency. This is comparable to the actual 57\% decrease of the resonance frequency. The discrepancy is probably due to spatially nonuniform hydration across the large drum head (18-inch diameter).

\section*{Discussion}
\noindent
The technique we present is nearly model-free in the sense that it is independent of and does not require the material's mechanical properties.
Although our method assumes the material is incompressible, extending the method to accommodate compressibility adds a relative small error in the order of $\mu/\lambda$, where $\lambda$ is the first Lam\'e constant (Supplementary Note S2 and Fig. S2).
The plastic cling film we tested is actually a compressible material with an initial Poisson's ratio of $\sim 0.32$ and had significant plastic deformation.
The technique is primarily for elastic materials, but it can be applied to weakly viscoelastic materials~\cite{Bercoff2004}. However, highly viscoelastic materials with frequency-dependent mechanical properties call for more involved curve fitting with parameters including the viscosity~\cite{DeRooij2016}.

Although we have validated the technique for relatively uniform stress, it is easily applicable to more complex structures with spatially varying stress and strain fields by measuring local velocity profiles, as illustrated in Fig.~\ref{fig:figure1}c. The spatial resolution of the mapping is approximately in the order of one wavelength. When $v$ = 10 m/s, for example, the resolution is $\sim$ 5 mm for $f$ = 2-10 kHz.
It is relatively straightforward to extend our method to thin structures in contact with fluids or gel-like matters either one side or both sides, such as dura mater of the brain after craniotomy~\cite{Hartmann2021} and the cornea~\cite{Antoine2019OE} as well as blood vessel walls~\cite{li2017guided}. Our work is expected to pave the way to practical applications.

\vspace{5 mm}

\vspace{5 mm}
\section*{Methods}
\subsubsection*{Theoretical model}

\setlength{\belowdisplayskip}{12pt} \setlength{\belowdisplayshortskip}{12pt}
\setlength{\abovedisplayskip}{12pt} \setlength{\abovedisplayshortskip}{12pt}

We consider an elastic wave polarised in the $x_i-x_3$ plane, propagating along the $x_i$ axis ($i=1$ or $2$), with attenuation along the $x_3$ axis: $\mathbf{u} = \mathbf{u}_0 \exp{(-skx_3)} \exp{[ik(x_i - v_i t)]}$.
Inserting $\mathbf{u}$ into $\mathcal A^0_{pjqk}\partial^2 u_k/ \partial x_p \partial x_q - \partial \hat{\Bar{p}} / \partial x_j = \rho \partial^2 u_j / \partial t^2$ and eliminating $\hat{\Bar{p}}$ with the incompressibility $\nabla \cdot \mathbf{u} = 0$, we get the secular equation
\begin{equation}\label{eq:secular}
    \gamma_i  s^4 - (2\beta_i  - \rho v_i^2)s^2 + \alpha_i  - \rho v_i^2 = 0,
\end{equation}
where $\alpha_i  = \mathcal A^0_{i3i3}$, $2\beta_i  = \mathcal A^0_{iiii} + \mathcal A^0_{3333} - 2\mathcal A^0_{ii33} - 2 \mathcal A^0_{3ii3}$, and $\gamma_i  = \mathcal A^0_{3i3i}$.
With the stress-free boundary condition at $x_3 = \pm h$, we arrive at the dispersion equation for the Lamb wave~\cite{ogden1993effect}
\begin{equation} \label{dispersion-general}
\left(\frac{\tanh s_1 k_1h}{\tanh s_2 k_1h} \right)^{\pm 1}=  \dfrac{s_2(s_1^2+1)^2}{s_1(s_2^2+1))^2},
\end{equation}
where the exponent is $+1$ for symmetric modes and $-1$ for anti-symmetric modes, and $s_1^2$, $s_2^2$ are the roots of Eq.~\eqref{eq:secular}.

To capture the acousto-elastic effect induced by a moderate strain, we consider the strain energy of isotropic incompressible third-order elasticity~\cite{li2020ultrasonic,destrade2010third}. The elastic moduli $\mathcal A^0_{ijml}$ are given by Equation (9) in Li et al~\cite{li2020ultrasonic},  which result in  $2\beta_i = (\alpha_i + \gamma_i)$.
Substituting into Eq.~\eqref{eq:secular} we get $s_1^2 = (\alpha_i  - \rho v_i^2) / \gamma_i $ and $s_2^2=1$, and the dispersion equation Eq.~\eqref{dispersion-general} becomes Eq.~\eqref{eq:dispersion}.

The basic idea of our acousto-elastic imaging technique is to deduce the stress and strain  with $\alpha_i$ and $\gamma_i$ from the exact formulas
\begin{equation}\label{eq:stress}
    \sigma_i = \alpha_i - \gamma_i, \qquad \
    \lambda_i^2 / \lambda_3^2= \alpha_i / \gamma_i.
\end{equation}
To show these formulas, it suffices to recall that, in general~\cite{Ogden2007},
\begin{equation}\label{A0ijij}
    \mathcal A^0_{jkjk} = \frac{\sigma_j - \sigma_k}{\lambda_j^2 - \lambda_k^2} \lambda_j^2, \quad (\text{for} \;\; j \neq k \text{ and } \lambda_j \neq \lambda_k).
\end{equation}
Making use of Eq.~\eqref{A0ijij}, we get $\alpha_i - \gamma_i = \mathcal A^0_{i3i3} - \mathcal A^0_{3i3i} = \sigma_i - \sigma_3$, and $\alpha_i / \gamma_i = \mathcal A^0_{i3i3} / \mathcal A^0_{3i3i} = \lambda_i^2 / \lambda_3^2$, which leads to Eq.~\eqref{eq:stress} by taking $\sigma_3 = 0$ (thin membrane).

Expanding \eqref{dispersion-general} in the low frequency limit of $f = 0$ (or $kh = 0$), we get $\rho v_{A0}^2 = \alpha_1 - \gamma_1$ and $\rho v_{S0}^2 = \alpha_1 + 3\gamma_1$ for the A0 and S0 modes, respectively.
On the other hand, in the high frequency limit of $f \to \infty$ (or $kh \to \infty$), we get $\rho v_{R}^2 = \alpha_1 - \eta_0^2\gamma_1$ for both A0 and S0 modes, where $\eta_0= 0.2956$ is the real root of the cubic $x^3 + x^2 + 3x - 1 = 0$ (Rayleigh surface wave limit).
The three limits are shown in Fig.~\ref{fig:figure2}d.

As a simple example, suppose an initially isotropic material is subject to a small stress.
In the limit of small deformation ($\lambda_i \approx 1)$, using Taylor expansion the three limits can be explicitly expressed as functions of the principal stresses
 \begin{equation}
     \begin{pmatrix} \rho \, v_{A0}^2(f=0) \\ \rho \, v_{S0}^2(f=0)\\ \rho \, v_{R}^2(f=\infty) \end{pmatrix} =
     \begin{pmatrix} 0 \\ 4 \mu \\ 0.91 \mu \end{pmatrix} +
     \begin{pmatrix} 1 & 0 & -1 \\ -2/3 & -2/3 & 4/3 \\ 0.62 & -0.15 & -0.47 \end{pmatrix}
     \begin{pmatrix} \sigma_1 \\ \sigma_2 \\ \sigma_3 \end{pmatrix}.
 \end{equation}
For thin structures, $\sigma_3 \approx 0$.
If two of these three limiting wave speeds are measured, the in-plane stresses ($\sigma_i$) can be calculated from this equation with $\sigma_3 = 0$.
In practice, the $kh \to \infty$ limit cannot be measured accurately because it corresponds to an extremely thick slab.
Similarly it is difficult to express the $S_0$ mode for ultra-thin films; in that case, the first equation for the $A_0$ mode nonetheless gives access to $\sigma_1-\sigma_3$ directly.

\subsubsection*{Experimental setup}
Our experimental setup (Supplementary Fig. S3) is based on a home-built, swept-source optical coherence tomography (OCT) system \cite{Antoine2019OE}.
This system offers an A-line rate of 43.2 kHz, axial resolution of $\sim 15$ $\mu$m (in the air) and transverse resolution of $\sim30$ $\mu$m, using a polygon swept laser with a tuning range of 80 nm and a centre wavelength of 1,280 nm.
The optical beam is scanned using a two-axis galvanometer scanner.
To excite Lamb waves in the film we used a contact probe driven by a vibrating PZT piezoelectric transducer (Thorlabs, PA4CEW).
The plastic probe was 3D-printed with a spherical tip of $~2$ mm in diameter.
A small force ($\sim 20$ mN) was applied to the probe to keep it in contact with the sample.
The optical beam scan was synchronized with the probe vibration to operate in an M-B scan mode (see Supplementary Note 3 and Fig. S10). Their scanner axes were aligned to the principal transverse axes ($x_i$).
The frequency of the vibration was step-tuned from 2 to 20 kHz with an interval of 2 kHz.
At each frequency, the amplitudes and phases of the vibrations were acquired at 96 transverse locations.
The vertical displacement near the probe contact point at the sample was in the order of 100 nm in the frequency range.
To measure this small vibration, we used the phase change in the interference signal of the OCT~\cite{Antoine2019OE}.

To measure the wavenumber $k$ and thus the wave speed $v_i = 2\pi f/k$ for a given frequency, the surface displacement was Fourier-transformed from the spatial domain to the wavenumber domain.
The wavenumber was obtained by identifying the peak that corresponds to the A0 mode (Supplementary Fig. S5).
The standard deviation error in the wavenumber measurement is estimated to be about $0.1\%$ (Supplementary Note 4).

For the experiments of the rubber sheet and cling film, the sample was clamped along its two short edges and one clamp was pulled horizontally by a cord connected to 1 to 6 weights ($m = $20 g each) to apply a uniaxial tension $\sigma_1$ with varying magnitudes.
The Cauchy stress applied to the film is $\sigma_1 = \lambda_1 \, N m \, g/(2\, h_0 \, W_0)$, where $W_0$ and $h_0$ are the initial width and half-thickness of the sample, respectively, $g = 9.8$ m/s$^2$ is the acceleration of gravity and $N$ is the number of the weights.
When changing the stress state, the results obtained during loading and unloading were averaged to minimise the effect of hysteresis.

For each stress state, the current thickness $2h$ of the rubber film was measured from the OCT image. The uniaxial stretch ratio was then determined by $\lambda_1 = \lambda_3^{-2} = (h_0/h)^2$. The measured stretch ratio agreed well with that obtained by the deformation of the grids drew on the surface of the rubber film.

\subsubsection*{Materials}
The rubber sheet has mass density $\rho \simeq$ 1,070 kg/m$^3$ and refractive index $n \simeq 1.4$.
The initial dimension was $2h_0 \simeq 0.5$ mm, $W_0 = 16$ mm, and $H_0 = 40$ mm. The lateral size was large enough to avoid wave reflections at the edges.
The rubber sample was prepared from Ecoflex$^\text{TM}$ 5 material (Smooth-On Inc) by mixing the Ecoflex 1A and 1B at 1:1 ratio by weight.
The mixture was poured into a mold and cured at room temperature overnight. Then
the material was post-cured in an oven at 80$^\circ$C for 2 hours.
For mechanical testing, we cut out a small piece ( $0.5 \times 5 \times 18$ mm$^3$) and performed a tensile test with a uniaxial tensile testing machine (eXpert 4000 Micro Tester, Admet, Norwood, USA).
Figure~\ref{fig:figure3}e shows the resulting stress-stretch curve.
Applying a linear fitting to the initial stage of the curve (stretch ratio < 1.07) we find that the initial shear modulus (one third of the Young modulus) is approximately 180 kPa.

We used a common home-use \emph{cling film} (plastic wrap) made of $100\%$ polyethylene with $\rho \simeq 930$ kg/m$^3$.
The typical thickness of cling films ranges from  8 to 13 $\mu$m.
Here we  used Brillouin microscopy \cite{Scarcelli2015} to measure the thickness of our film to be $11.7 \pm 0.3 \mu$m.
This is close to the axial resolution of the OCT system, and so we could not track $h$ with the deformation.

The \emph{bodhr\'an} instrument was purchased from Hobgoblin Music, MN, USA. The OCT measurement was performed on the intact instrument.
On separate direct measurements after removing the skin from the frame, we found $h_0 = 360 \pm 30$ $\mu$m and $\rho = 831 \pm 65$ kg/m$^3$ in the dry condition. After hydration, the density is expected to increase to $\sim$ 1000 kg/m$^3$.
To characterise the fundamental resonance frequencies of the instrument in the dry and damp conditions, the centre of the drumhead was beaten every 10 seconds while recording the sound with a cellphone 10 cm away from the drumhead, using the Google Science Journal App.

\section*{Acknowledgements}

This study was supported by grants P41-EB015903, R01-EB027653, DP1-EB024242 from the National Institutes of Health (USA) for G.Y.L and S.H.Y,  and by the 111 Project for International Collaboration No. B21034 (Chinese Government, PR China), a grant from the Seagull Program (Zhejiang Province, PR China) for M.D and a grant from the European Commission - Horizon 2020 / H2020 - Shift2Rail for A.L.G.
The authors thank Drs Amira Eltony and Xu Feng for help with the measurements, and Pasquale Ciarletta, Niall Colgan and Giuseppe Zurlo for valuable feedback.

\section*{Author contributions statement}

G.Y.L., A.G., and M.D. designed the study. G.Y.L., A.G., and M.D. developed the theoretical model. G.Y.L. conducted the experiments. G.Y.L. and S.H.Y. analyzed the results. All authors wrote and reviewed the manuscript.

\end{document}



\title{Supporting Information \\ for \\ \textit{Non-destructive mapping of stress, strain and stiffness of thin elastically deformed materials}}

\date{\today}

\author{Guo-Yang Li}
\affiliation{Harvard Medical School and Wellman Center for Photomedicine, Massachusetts General Hospital, Boston, MA, USA;}%

\author{Artur L Gower}
\affiliation{Department of Mechanical Engineering, University of Sheffield, Sheffield, United Kingdom;}%

\author{Michel Destrade}
\affiliation{School of Mathematical and Statistical Sciences, NUI Galway, Galway, Ireland;\\
 Key Laboratory of Soft Machines and Smart Devices of Zhejiang Province, Department of Engineering Mechanics, Zhejiang University, Hangzhou 310027, PR China;}%

\author{Seok-Hyun Yun}
\affiliation{Harvard Medical School and Wellman Center for Photomedicine, Massachusetts General Hospital, Boston, MA, USA.}%

\maketitle

\renewcommand{\thepage}{S\arabic{page}}
\renewcommand{\thetable}{S\arabic{table}}
\renewcommand{\thefigure}{S\arabic{figure}}
\renewcommand{\theequation}{S.\arabic{equation}} 

\textbf{Supplementary Results}:

Supplementary notes 1-4

Figs. S1-S10

Supplementary movies 1-4

\newpage
\section*{Supplementary Note 1: Sensitivity analysis}

Here we demonstrate how sensitive our prediction of the stress $\sigma_1$ is to small errors in our measurements.
The analysis  focuses on the anti-symmetric mode A0 (the results for the S0 mode are similar).

\begin{figure}[bh]
  \includegraphics[height=0.4\linewidth]{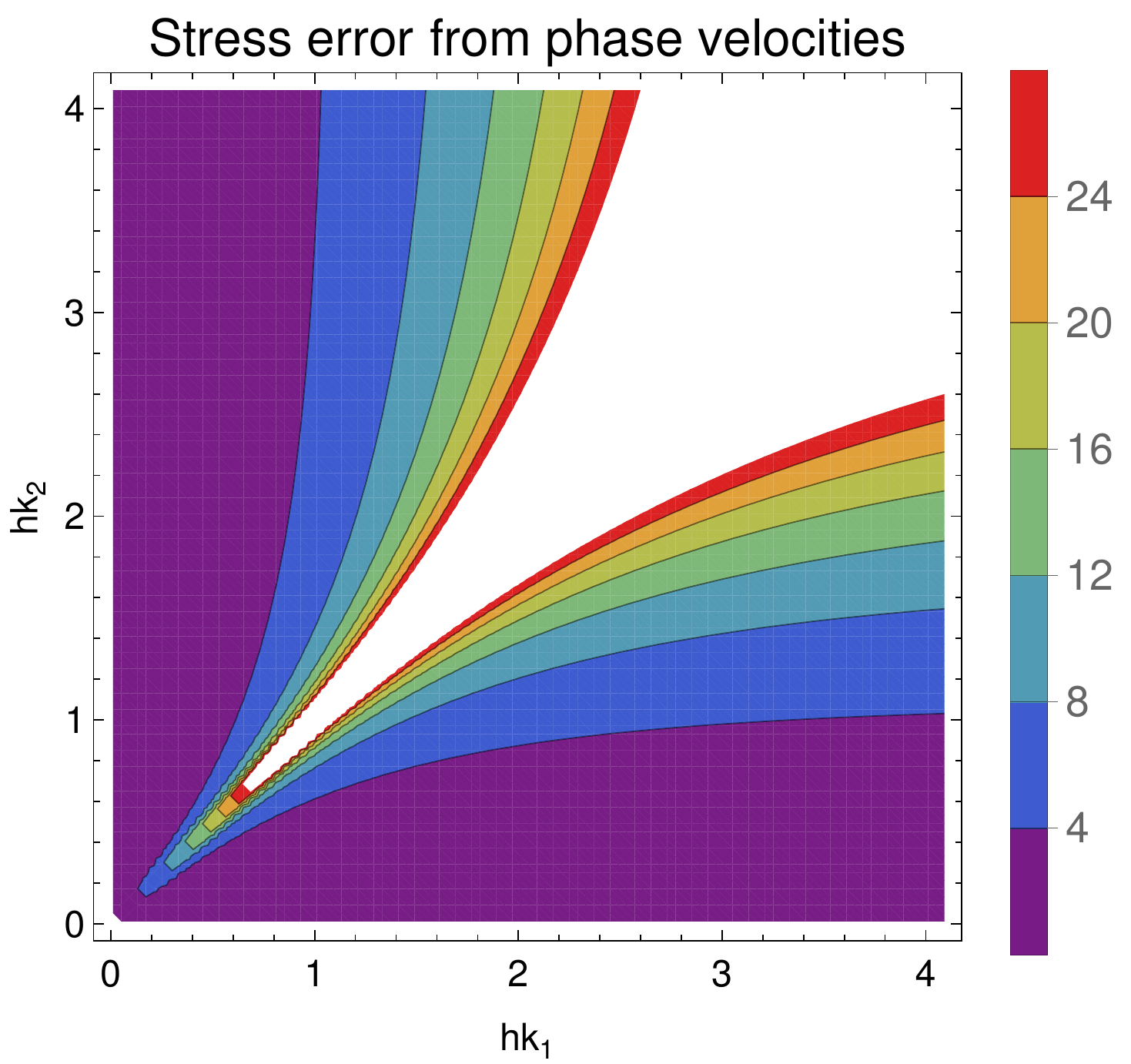}
  \includegraphics[height=0.4\linewidth]{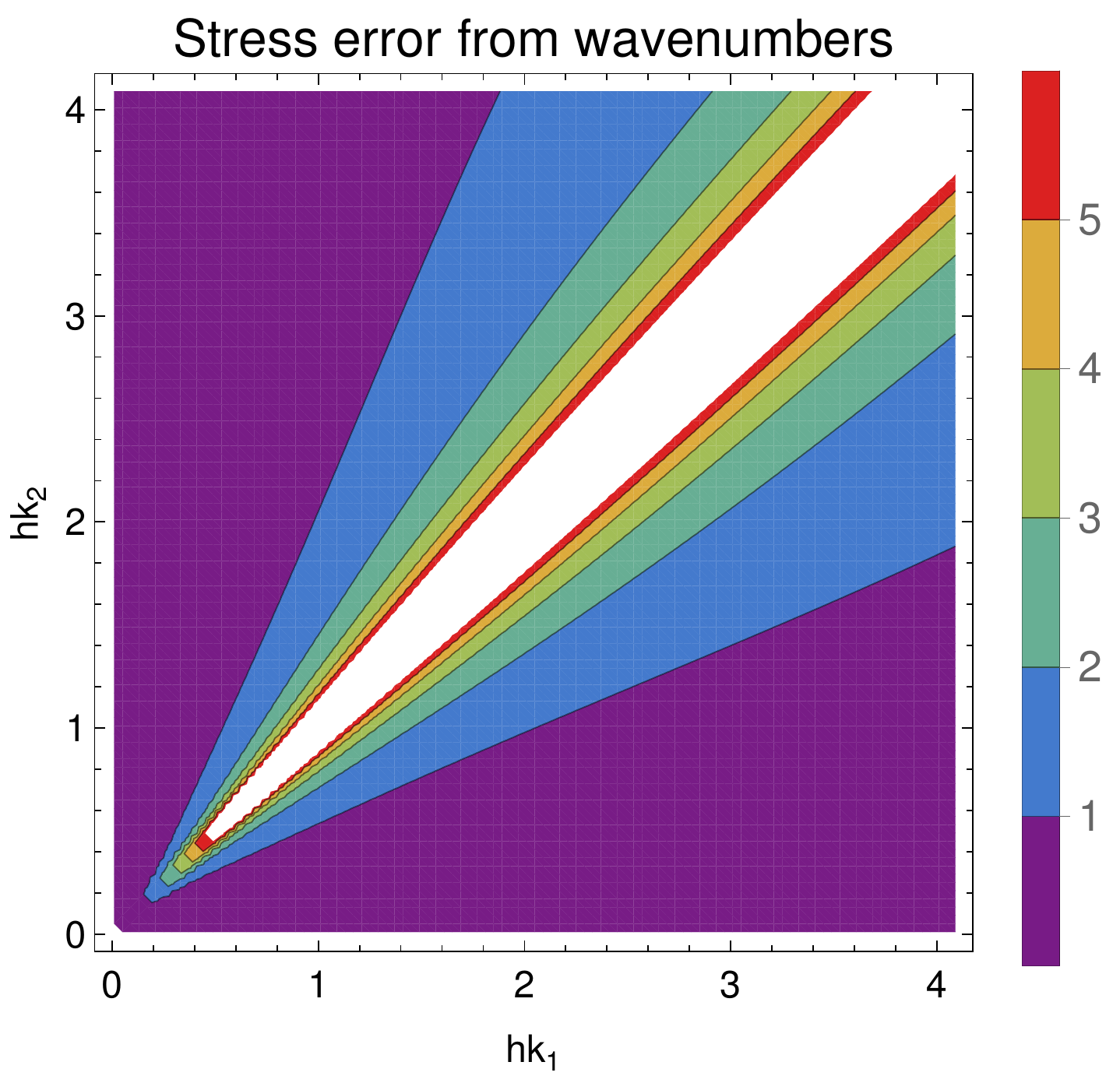}
  \vspace{-0.7cm}

 \hspace{0.02\linewidth} \textbf{a} \hspace{0.4\linewidth}  \textbf{b} \hspace{0.44\linewidth}

\vspace{0.2cm}

 \includegraphics[height=0.4\linewidth]{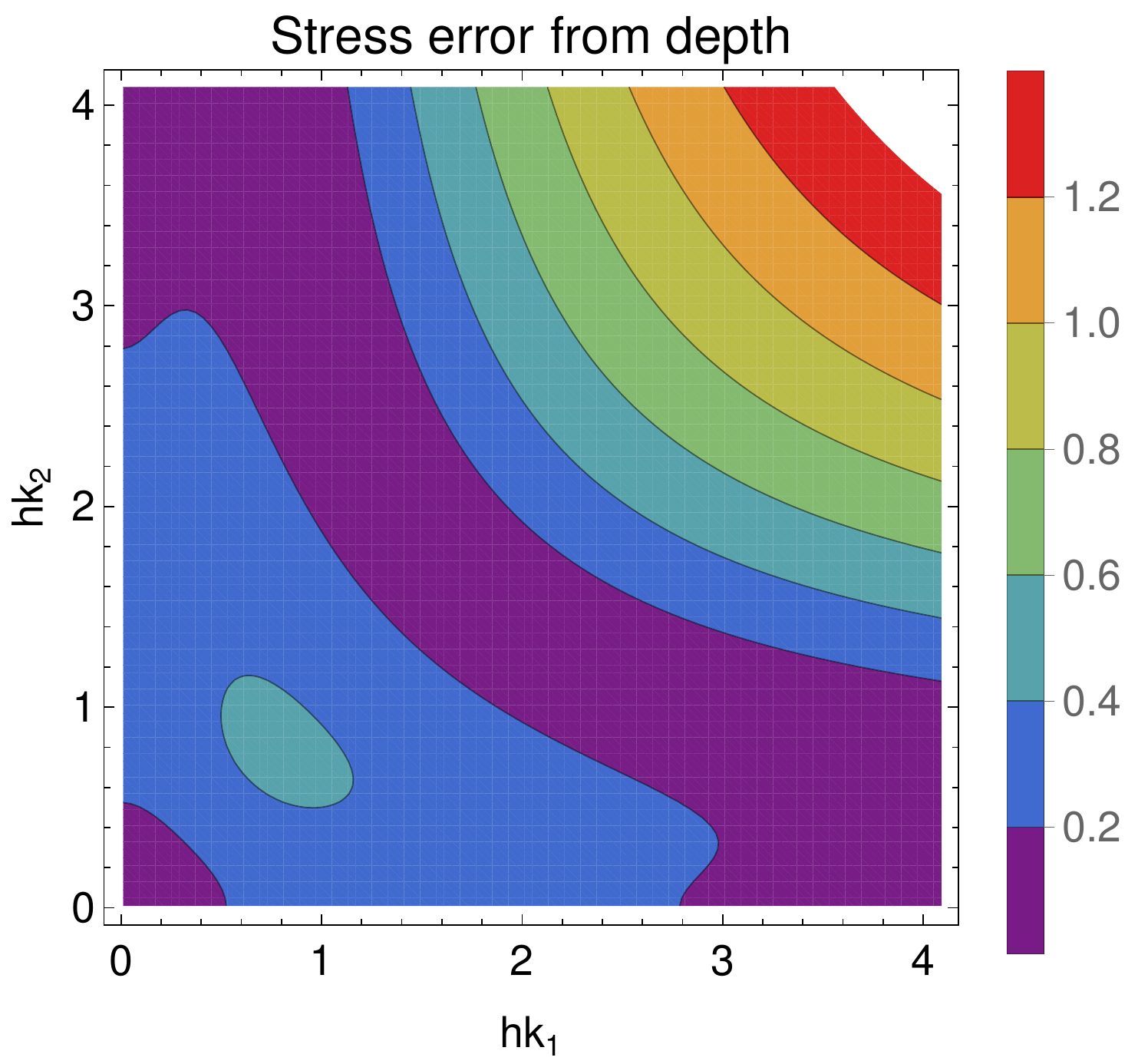}
 \includegraphics[height=0.4\linewidth]{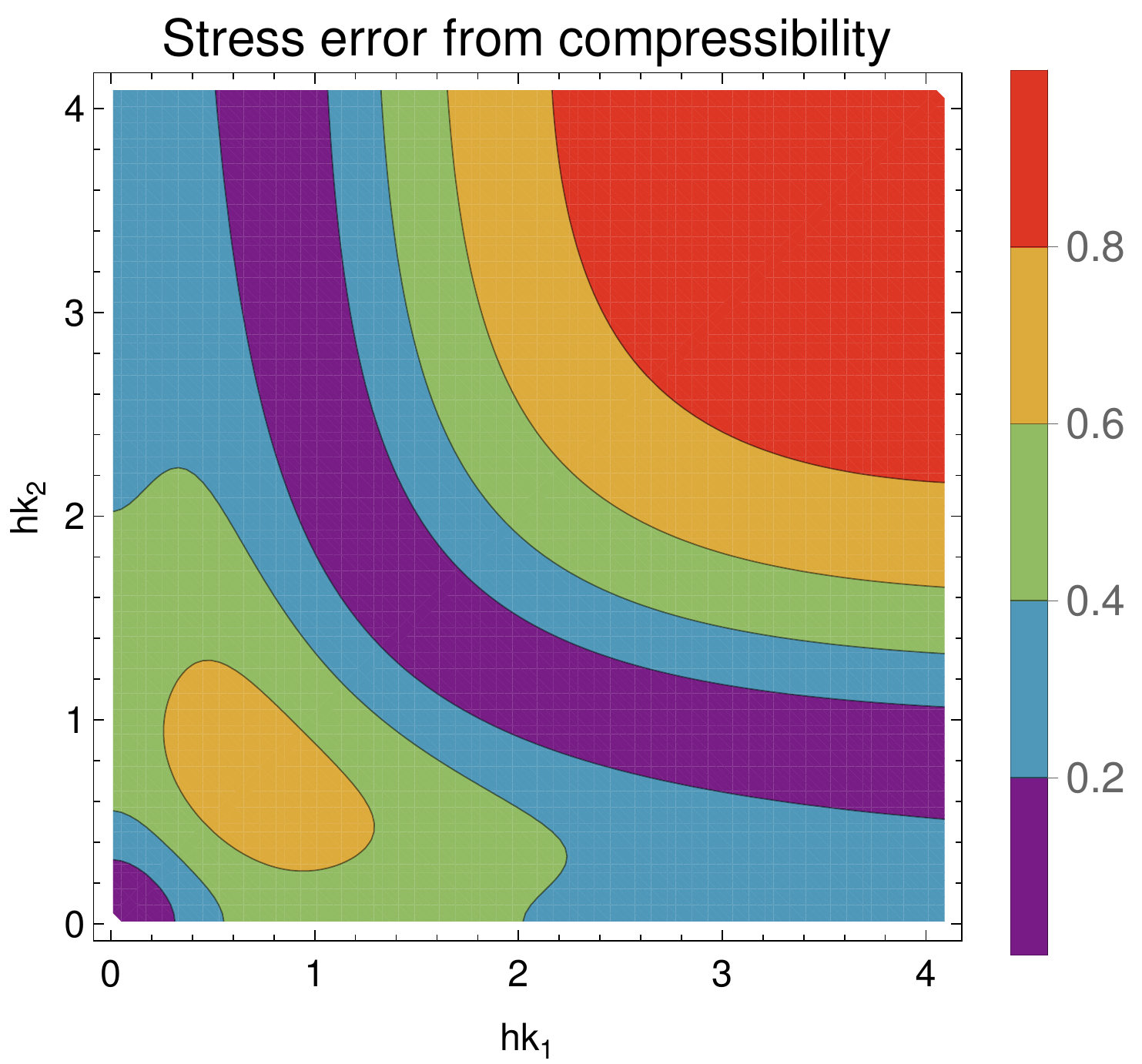}
  \vspace{-0.7cm}

 \hspace{0.02\linewidth} \textbf{c} \hspace{0.4\linewidth}  \textbf{d} \hspace{0.44\linewidth}
 \caption{ \textbf{Sensitivity in the prediction of stress to different sources of error.}
 \textbf{a}, Effect of the relative errors in $v_1$ and $v_2$.
 \textbf{b}, Effect of the errors in estimating $h k_1$ and $h k_2$.
 \textbf{c}, Effect of the errors in estimating the depth $h$.
 \textbf{d} The values in this plot times $\mu /\lambda$ would be the error in the stress. For example, if $\mu/\lambda = 1$, and $\mu = 0.02$ GPa, then we would expect an error of 0.016 GPa in the stress when using $h k_1 \approx h k_2 \approx 4$.
 In all the plots, it is assumed that the phase speeds $v_1$ and $v_2$ of an anti-symmetric Lamb wave are measured at the wavenumbers $h k_1$ and $h k_2$, respectively. The white regions have errors larger than the values shown in the accompanying legend. }
 \label{fig:sensitivity-analysis}
\end{figure}

Let $v_1$ and $v_2$~\cite{Note1} denote the A0 wave speeds and $k_1$ and $k_2$ the corresponding wavenumbers, respectively. We consider wave speeds along $x_1$ principal axis, as the results for the $x_2$ axis are identical. That is, the $v_1$ and $v_2$ here are not the phase speeds along the $x_1$ and $x_2$ axis as discussed in the paper.
We define $\eta_{1} = \sqrt{(\alpha_1 - \rho v_1^2) / \gamma_1}$ and $\eta_{2} = \sqrt{(\alpha_1 - \rho v_2^2) / \gamma_1}$. Using $\sigma_1 = \alpha_1 - \gamma_1$ we obtain
\begin{equation} \label{eqn:predict-stress}
  \sigma_1 = \tfrac{1}{2}(\rho v_1^2 + \rho v_2^2) + \tfrac{1}{2}(\rho v_1^2 - \rho v_2^2)F, \qquad \text{where} \qquad F = \frac{\eta_1^2 + \eta_2^2 - 2}{\eta_1^2 -\eta_2^2}.
\end{equation}

There are two main sources of errors in using this equation to predict the stress. The first is the measurement error in the square of the wave speeds, $\delta v_1^2$ and $\delta v_2^2$, and the second is the measurement error in the non-dimensional wavenumber parameters, $\delta(k_1 h)$ and $\delta (k_2 h)$. We investigate these errors separately.

First we assume there is an error only in measuring the wave speeds, expressed as $\rho \delta v_1^2$ and $ \rho \delta v_2^2$, so that the error in the stress $\delta \sigma_1$, according to Eq.~\eqref{eqn:predict-stress}, becomes
\begin{equation} \label{eqn:predict-stress-speed}
  \delta\sigma_1 = \tfrac{1}{2}\rho \delta v_1^2(1+ F) + \tfrac{1}{2}\rho \delta v_2^2( 1 - F)
  \qquad \text{so that} \qquad
   \frac{|\delta\sigma_1|}{\rho|\delta v^2|} \leq  \tfrac{1}{2}|1+ F| + \tfrac{1}{2}|1 - F|.
\end{equation}
Here we have assumed that $\delta v_1$ and $\delta v_2$ are random and uncorrelated, and used $|  \delta v^2|$ to represent the maximum error in the squared velocities.
This stress error is displayed in Fig.~\ref{fig:sensitivity-analysis}a.

Next, we assume that there is an error only in the wavenumber parameters. We can write
\begin{equation} \label{eqn:predict-stress-eta}
  \delta\sigma_1  = \tfrac{1}{2}(\rho v_1^2 - \rho v_2^2)\left[\frac{\partial F}{\partial \eta_1} \delta \eta_1 + \frac{\partial F}{\partial \eta_2} \delta \eta_2 \right].
\end{equation}
Using the exact relation $\rho v_1^2 - \rho v_2^2 = \gamma_1 (\eta_2^2 - \eta_1^2)$ we obtain
\begin{equation} \label{eqn:predict-stress-eta-2}
  \delta \sigma_1 = \gamma_1 \frac{\delta \eta_2^2 ( 1- \eta_1^2) - \delta \eta_1^2 (1- \eta_2^2)}{\eta_1^2 - \eta_2^2}.
\end{equation}
For third-order elasticity, we find $\gamma_1 \approx \mu$. Any error committed when calculating $\eta_1$ and $\eta_2$ result from the error in $k_1 h$ and $k_2 h$. We can then write
\begin{equation} \label{eqns:eta-errors}
 \delta \eta_1^2 = \frac{\partial \eta_1^2}{\partial (k_1h)} \delta (k_1 h) \quad \text{and} \quad
 \delta \eta_2^2 = \frac{\partial \eta_2^2}{\partial (k_2h)} \delta(k_2 h).
\end{equation}
There are two possible ways to commit the errors $\delta(k_1 h)$ and $\delta (k_2 h)$. The first is in the miscalculation of the wavenumbers $k_1$ and $k_2$. This results in $\delta(k_1 h) = h \delta k_1$ and $\delta(k_2 h) = h \delta k_2$. Assuming the errors are uncorrelated, we obtain
\begin{equation} \label{eqn:wavenumber-error}
 \frac{|\delta \sigma_1 |}{\mu |h \delta k|} \leq \left|\frac{\partial \eta_1^2}{\partial (k_1h)}\right| \frac{|1- \eta_1^2|}{|\eta_1^2 - \eta_2^2|} + \left|\frac{\partial \eta_2^2}{\partial (k_2h)}\right| \frac{|1- \eta_2^2|}{|\eta_1^2 - \eta_2^2|}.
\end{equation}
Note that typically $\mu > |\sigma_1|$. Therefore, the right hand-side is smaller than the relative error of the stress. The magnitude of the right hand-side term is plotted in Fig.~\ref{fig:sensitivity-analysis}b for a range of $\eta_1$ and $\eta_2$.

The second way to commit the errors $\delta(k_1 h)$ and $\delta (k_2 h)$ is due to miscalculation of the half thickness $h$. This results in $\delta(h k_1) =   k_1 \delta h$ and $\delta(h k_2) = k_2 \delta h$, from which we obtain
\begin{equation} \label{eqn:depth-error}
  \frac{|\delta \sigma_1|}{\mu|\delta h / h|} \leq  \frac{1}{|\eta_1^2 - \eta_2^2|}\left|\frac{\partial \eta_1^2}{\partial (k_1h)} k_1 h ( 1- \eta_1^2) - \frac{\partial \eta_2^2}{\partial (k_2h)} k_2 h (1- \eta_2^2) \right|.
\end{equation}
The magnitude of the right hand-side term is plotted in Fig.~\ref{fig:sensitivity-analysis}c.

\newpage
\section*{Supplementary Note 2: Material compressibility}

Our method requires that the material be incompressible.
Here we calculate errors for nearly incompressible materials.

\begin{figure}[ht!]
  \includegraphics[height=0.25\linewidth]{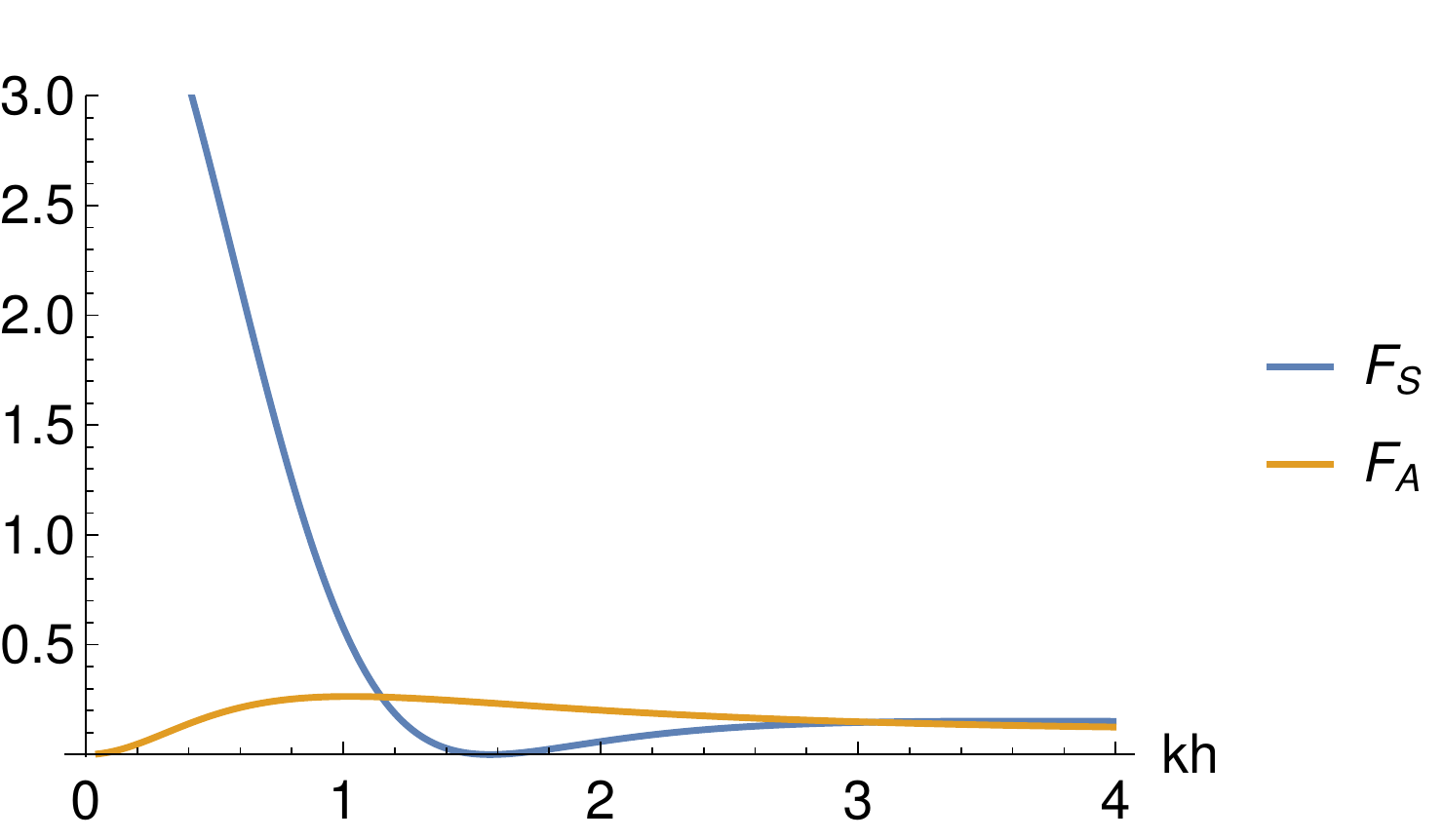}
  \caption{\textbf{Variation of the terms $F_A$ and $F_S$ in~\eqref{eqns:compressible-terms} with the wavenumber $kh$.}
  When $F_S$ or $F_A$ is small, the wave speed is insensitive to the material compressibility.}
  \label{fig:compressibility-terms}
\end{figure}

Equations governing Lamb waves in compressible solids have been given by Ogden and Roxburgh\cite{ogden1993effect}.
As we have done for the incompressible case, we specialise to third-order elasticity.
For small compressibility, $\mu / \lambda \ll 1$, where $\mu$ and $\lambda$ are Lam\'e constants of linear elasticity. We follow a method used by Shams \textit{et al}.~\cite{destrade2010third} using a Taylor expansion in terms of $\mu / \lambda$.
For small strain up to third-order elasticity~\cite{Li2020JASA}, we obtain
\begin{align} \notag
  & \rho v_A^2 = \gamma \eta_A + \alpha + \rho \delta v_A^2, \qquad  \frac{\rho \delta v_A^2}{\mu} = \frac{2 \mu}{\lambda} F_A(kh),
  \\ \label{eqns:compressible-terms}
  & \rho v_S^2 = \alpha - \gamma \eta_S +  + \rho \delta v_S^2, \qquad  \frac{\rho \delta v_S^2}{\mu} = \frac{2 \mu}{\lambda} F_S(kh),
\end{align}
for the anti-symmetric and symmetric modes, respectively. The terms $F_A(kh)$ and $F_S(kh)$ depend only on $kh$. For example
\[
F_A(kh) = \frac{(S_2 -2 kh) \eta_A^2 (1 - \eta_A^2) (1 + \eta_A^2)^2}{S (-4 C + kh S) - 2 ((5 + 3 C_2) kh - 2 S_2) \eta_A^2 +  6 S (2 C + kh S) \eta_A^4 + 4 kh S^2 \eta_A^6 + kh S^2 \eta_A^8}.
\]
where $S = \sinh(kh)$, $C = \cosh(kh)$,  $S_2 = \sinh(2kh)$, and $C_2 = \cosh(2kh)$. Both $F_A$ and $F_S$ are plotted in Fig.~\ref{fig:compressibility-terms}.

To investigate the error induced by the nearly incompressible materials, we use this and ~\eqref{eqn:predict-stress-speed} to arrive at
\[
\frac{|\delta\sigma_1|}{\mu} \frac{\lambda}{\mu} = \left | F_A(k_1h)( 1+ F) + F_A(k_2h) ( 1 - F) \right|.
\]
When ${\mu}/{\lambda} \ll 1$ this error is always small.
However, the error can be significant for considerably compressible materials, such as steel~\cite{Li2020JASA} with $\mu / \lambda \approx 0.8$. The numerical values of the right hand-side term of the above equation are plotted in Fig.~\ref{fig:sensitivity-analysis}d.

\newpage
\begin{figure}[ht!]
  \includegraphics{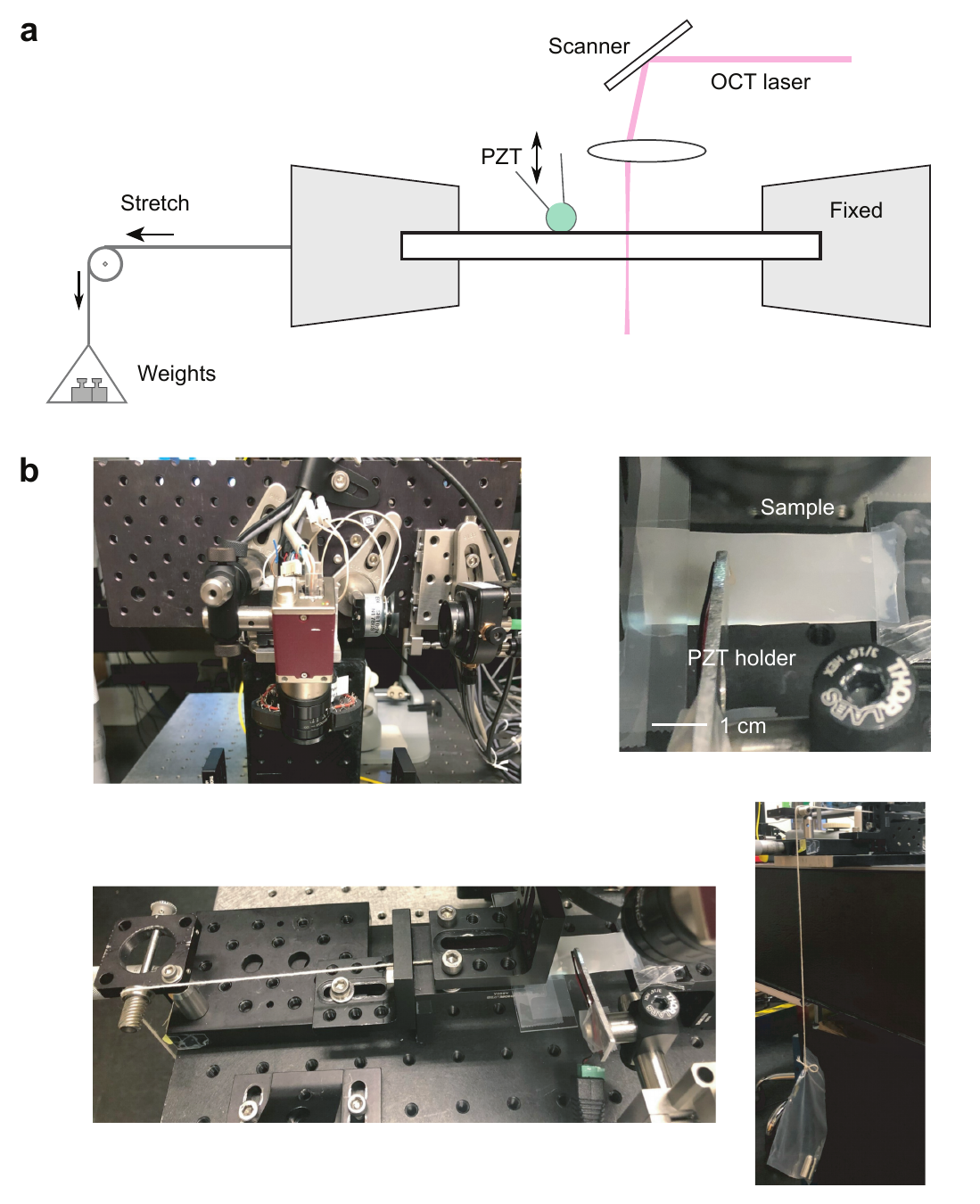}
  \caption{\textbf{Experimental setup.}
  \textbf{a}, Illustration of the experimental set up,
  \textbf{b}, Photographs of the sample and the probe.}
  \label{fig:experimentsetup}
\end{figure}

\newpage
\begin{figure}[ht!]
  \includegraphics[width=0.3\textwidth]{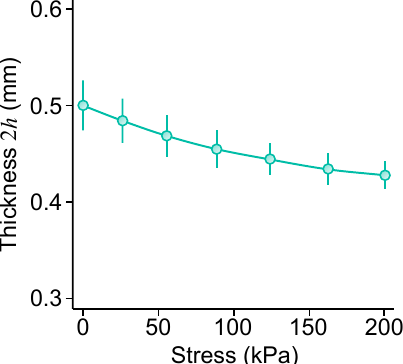}
  \caption{\textbf{Variation of the rubber sheet thickness (as tracked by OCT) with the stress, from $N = 0$ (stress-free) to $N = 6$ weights of 20 g each applied to create the stress. As expected, the thickness decreases as we increase the stress, due to the Poisson effect.}}
  \label{fig:thickness}
\end{figure}

\newpage
\begin{figure}[ht!]
  \includegraphics{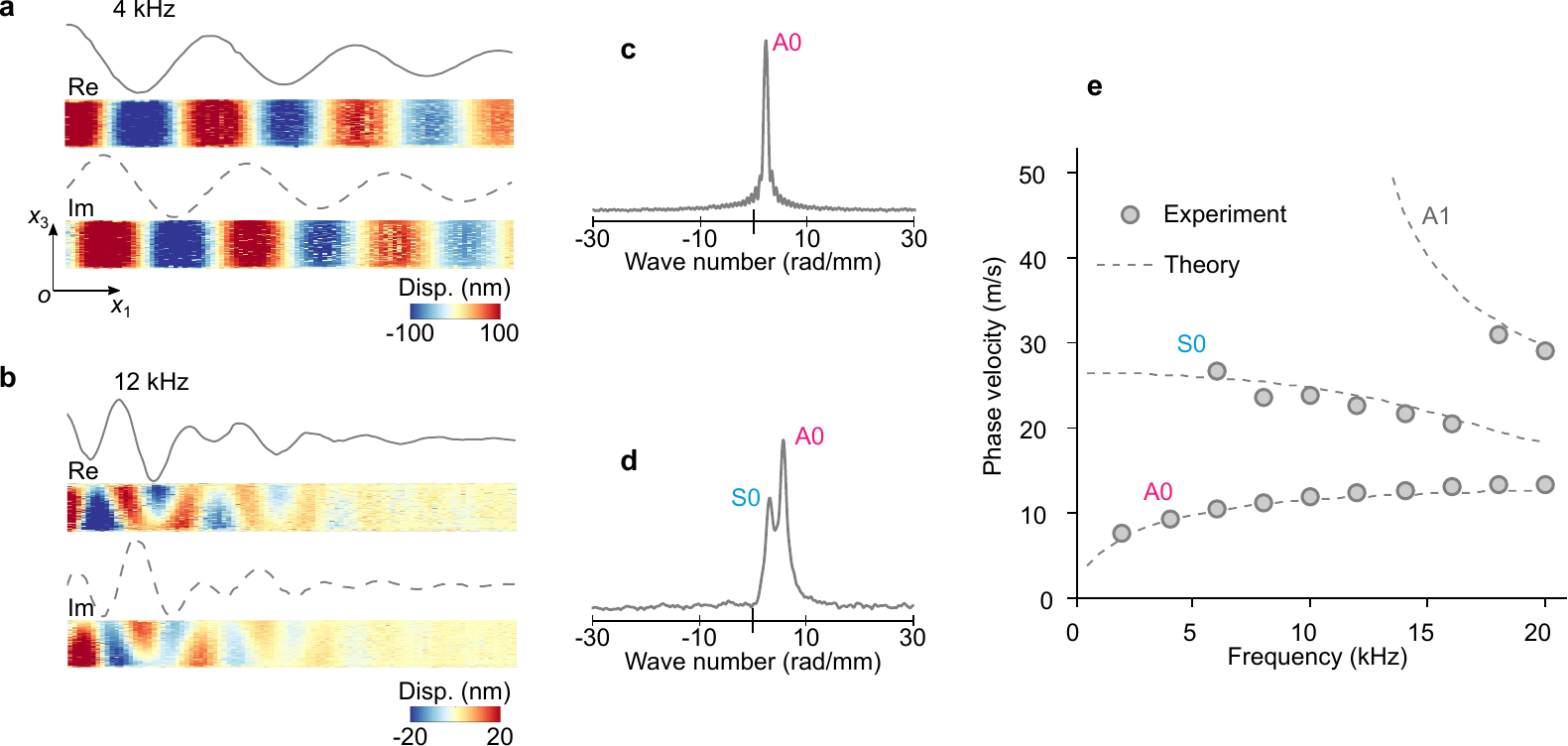}
    \caption{\textbf{Calculation of the wave speed by the Fourier transformation of the displacement profile.}
    \textbf{a}, The real and imagery parts of the displacement when the stimulus frequency $f$ is 4 kHz. The curves show the displacements extracted from the top surface of the sample.
    \textbf{b}, Displacement when $f$ is 12 kHz.
    \textbf{c} and \textbf{d}, Wavenumber domain data obtained by the Fourier transformation of the spatial domain data shown in \textbf{a} and \textbf{b}.
    Only one peak that corresponds to the A0 mode can be clearly identified from the curve in \textbf{c}, which suggests the A0 mode is predominantly excited at the stimulus frequency.
    However, two peaks that correspond to the A0 and S0 modes, can be identified in \textbf{d}.
    When the stimulus frequency is higher than 14 kHz, we find the high-order mode, A1 is also excited.
    \textbf{e}, The phase velocities of different waves modes measured when the sample is stress-free. The dashed curves are theoretical solutions computed using the shear modulus obtained from the tensile test ($\sim183$ kPa).
    Our sensitivity analysis suggests that the A0 mode in the low frequency range $f < 6$ kHz give the best sensitivity to the stress along the wave propagation direction.
    }
    \label{fig:speed measurement}
\end{figure}

\newpage
\begin{figure}[ht!]
  \includegraphics[width = 0.35\linewidth]{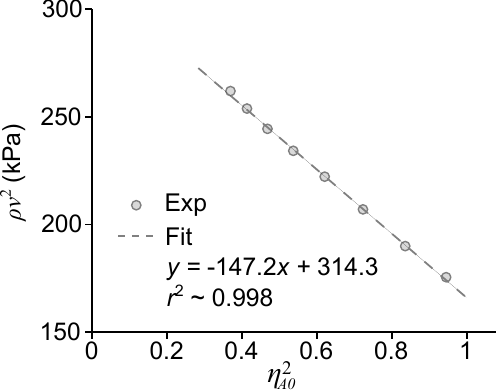}
  \caption{\textbf{A representative fitting ($N = 5$ weights applied) to deduce $\alpha$ and $\gamma$.}
  From the fitting we get $\alpha = 314$ kPa and $\gamma = 147$ kPa. Therefore, the stress $\sigma_1$ and stretch $\lambda_1$ are 167 kPa and 1.29, which agree well with the applied stress 162.1 kPa and stretch 1.31.}
  \label{fig:fitting}
\end{figure}

\newpage
\begin{figure}[ht!]
  \includegraphics[width = 0.5\linewidth]{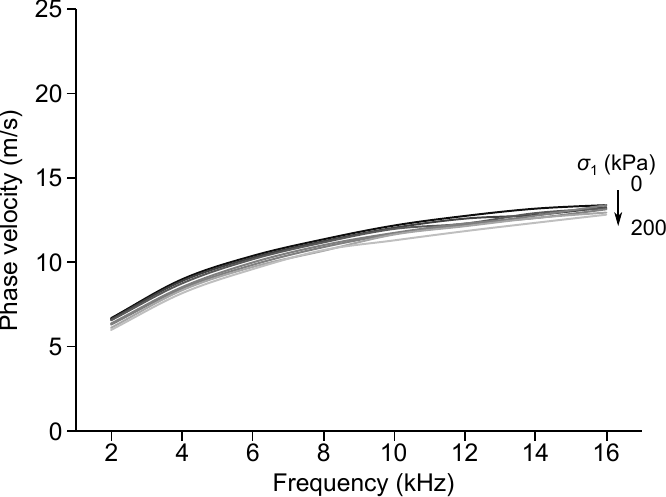}
  \caption{\textbf{Dispersion curves measured perpendicular to the uniaxial stress.}
  The phase velocity in the $x_2$ direction decreases as the tensile stress $\sigma_1$ is increased.}
  \label{fig:perpendicular}
\end{figure}

\newpage
\begin{figure}[ht!]
  \includegraphics{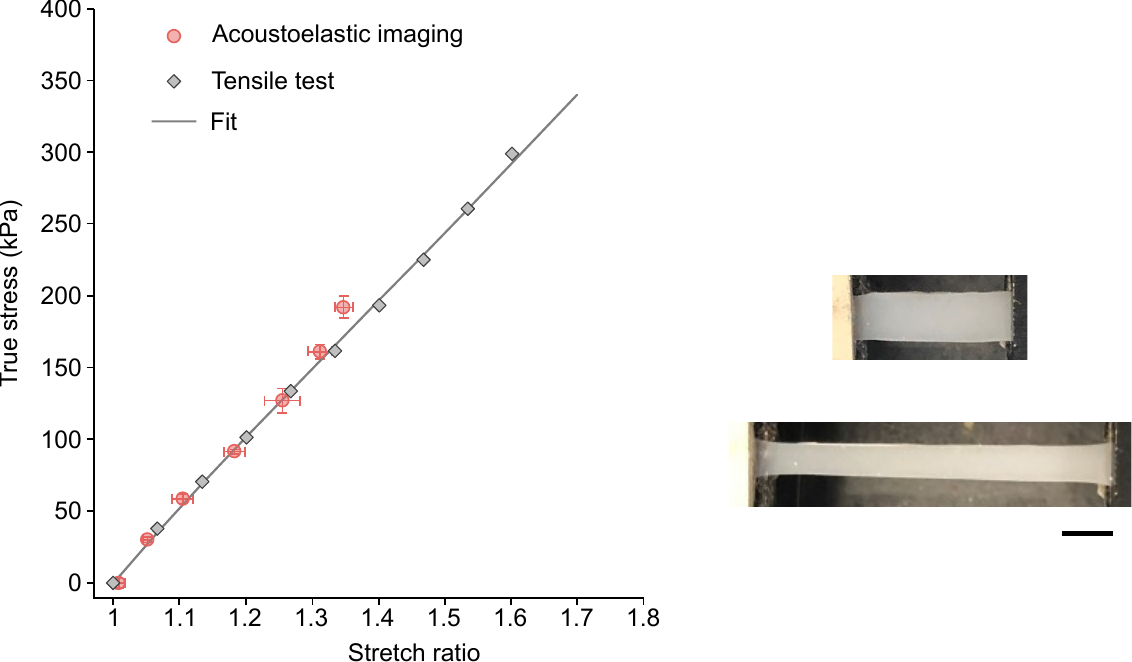}
  \caption{\textbf{Tensile test of the sample.}
  The shear modulus $\mu \approx 183$ kPa is obtained by fitting the initial stage (stretch ratio $<1.1$) of the tensile test data.
  To fit all the data we use the Mooney-Rivlin model $W = C_{10}(\lambda_1^2 + \lambda_2^2 + \lambda_3^2 - 3) + C_{01}(\lambda_1^2 \lambda_2^2 + \lambda_2^2 \lambda_3^2 + \lambda_3^2 \lambda_1^2 - 3)$, with $C_{10} = 49.2$ kPa and $C_{01} = 42.1$ kPa.
 The connections with the moduli of third-order elasticity are $\mu  = 2(C_{10} + C_{01})$ and $A = -8(C_{10} + 2C_{01})$, or here, $\mu = 183$, $A = -1,067$ kPa.
These material parameters are used in the main text to produce theoretical dispersion curves and confirm the match with the experimental data, although ultimately they are not needed for our stress measurement method through OCT imaging.
  We also plot the stress and strain obtained from acousto-elastic imaging.
  The good agreement with the independent tensile test verifies the validity of our method.
  The right panel shows photographs of the sample in the original and deformed conditions. Scale bar, 5 mm.}
  \label{fig:tensile test}
\end{figure}

\newpage
\begin{figure}[ht!]
  \includegraphics{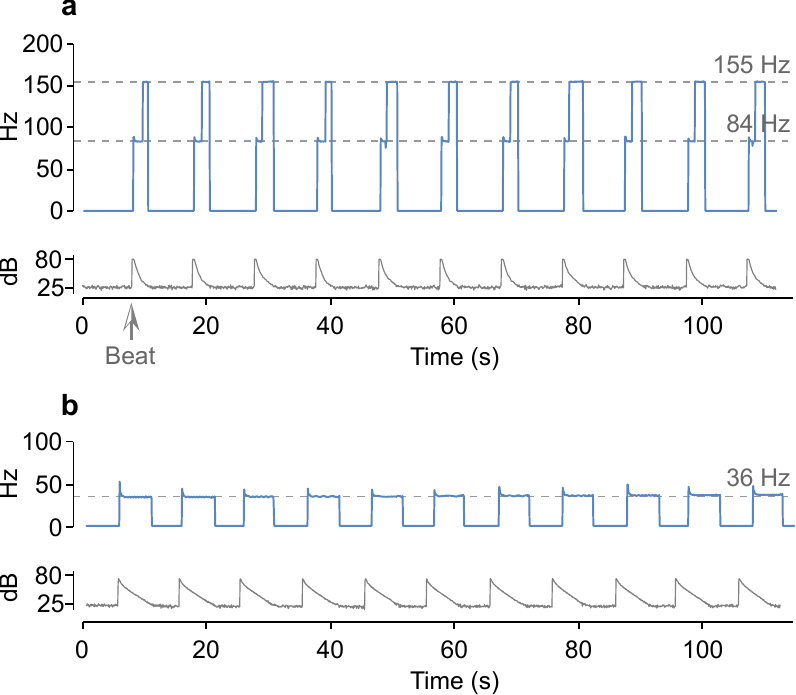}
  \caption{\textbf{Frequency change of \emph{bodhr\'an drum} due to humidification.}
  The drum was beaten every 10 s.
  \textbf{a}, Dry state. The fundamental and second harmonic frequencies are $\sim84$ Hz and $\sim 155$ Hz.
  \textbf{b}, Wet state. The fundamental frequency drops to $\sim 36$ Hz.}
  \label{fig:drum frequency}
\end{figure}

\newpage
\section*{Supplementary Note 3: M-B scan with optical coherence tomography}

\begin{figure}[hb!]
    \centering
    \includegraphics[width=.55\textwidth]{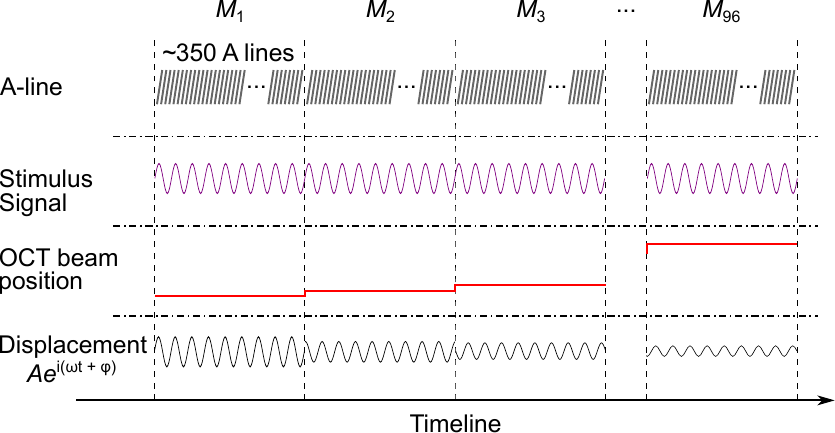}
    \caption{\textbf{Schematic of the M-B scan.}}
    \label{fig:figureS_MBScan}
\end{figure}

The optical coherence tomography (OCT) system was operated in a M-B scan mode, which is illustrated in Fig. \ref{fig:figureS_MBScan}.
The laser beam is scanned synchronously with the stimulus signal sent to the PZT.
At each lateral location, $\sim 350$ A lines (M scan) are acquired at a sampling rate of $\sim 43$ kHz.
A total 96 lateral locations are measured over a scan length of 5-12 mm (or $\sim$ 3 wavelengths of the A0 wave).
The vibration acquired from each M scan is Fourier transformed to obtain the amplitude $A$ and phase $\varphi$ in $u_3(t) = Ae^{i(\omega t + \varphi)}$.
Finally we obtain the real and imaginary parts of the displacement, $A\cos{(\omega t + \varphi)}$ and $A\sin{(\omega t + \varphi)}$.

\newpage
\section*{Supplementary Note 4: Sensitivity in measuring the wavenumber using OCT}

The standard deviation in the measurement of the vibration amplitude $A$ via a single A-line scan, denoted by $\delta A$, is given by optical signal-to-noise ratio ($\mathrm{SNR}$)\cite{Antoine_BOE2018}: $\delta A = \lambda_0 / (4\pi n_0 \sqrt{\mathrm{SNR}})$, where $\lambda_0$ is the optical wavelength ($\sim1280$ nm) and $n_0$ is the refractive index ($\sim1.4$).
At the surface of the sample, we typically get $\mathrm{SNR} \approx 40$ dB.
This sensitivity is improved by a factor $1/\sqrt{M}$ upon averaging of $M$ A-lines.
The elastic wave profile is obtained by measuring the displacement at $N$ locations along the propagation direction and then Fourier-transformed to determine its wavenumber $k$.
When the beam scan length covers $\sim3$ wavelengths, we find the standard deviation error of the wavenumber, denoted by $\delta k$, is  given by $\delta k / k \approx \delta A / (A \sqrt{MN})$.
With $\Delta A = 20$ nm (see Fig.~\ref{fig:speed measurement}b), $M = 350$, and $N = 96$, we obtain $\delta k / k \approx 0.1\%$. This SNR-induced error would be negligible for most applications.

%

\newpage
\vspace{10 mm}
\section*{The captions of movie files}

\vspace{10 mm}
\textbf{Supplementary movie 1}: Lamb wave propagation within the rubber sheet in the stress free state. Stimulus frequency, 6 kHz. Scale bar, 4 mm.

\vspace{10 mm}
\textbf{Supplementary movie 2}: Lamb wave propagation within the rubber sheet stretched along the horizontal direction (by 120 g weight). Stimulus frequency, 6 kHz. Scale bar, 4 mm.

\vspace{10 mm}
\textbf{Supplementary movie 3}: Lamb wave propagation within the drum head in a dry state. Stimulus frequency, 16 kHz. Scale bar, 8 mm.

\vspace{10 mm}
\textbf{Supplementary movie 4}: Lamb wave propagation within the drum head in a damp state. Stimulus frequency, 16 kHz. Scale bar, 8 mm.